\documentclass[a4paper,12pt,eqno]{article}
\usepackage{theorem}
\newtheorem{thm}{THEOREM}

\theoremheaderfont{\scshape}

\title{{\Large {\bf A Path Integral Approach \\ for Disordered Quantum Walks in One Dimension}
}}
\author{{\small Norio Konno} \\
{\small {\it Department of Applied Mathematics}} \\
{\small {\it Yokohama National University}} \\
{\small {\it Hodogaya-ku, Yokohama 240-8501, Japan}} \\
{\small {\it norio@mathlab.sci.ynu.ac.jp}}
}

\date{\empty}
\begin{document}
\maketitle

\par\noindent
\begin{small}
\par\noindent
{\bf Abstract}. The present letter gives a rigorous way from quantum to classical random walks by introducing an independent random fluctuation and then taking expectations based on a path integral approach.

\footnote[0]{
{\it PACS numbers: 03.67.Lx, 05.40.Fb}
}
\footnote[0]{
{\it Key words.} Quantum walks; Path integral; Qunatum algorithms; Classical random walks.

}

\end{small}

\setcounter{equation}{0}

\section{Introduction}

Recently the quantum walk (QW) is receiving significant attention, because it is a natural generalization of classical random walks to quantum systems, and because QWs could be useful in constructing efficient quantum algorithms. Here we focus on discrete-time QWs on the integers. QWs behave quite differently from classical walks. For the classical walk, the probability distribution is a binomial distribution with a variance that grows linearly with time. The probability distribution in a QW, by contrast, has a complicated and oscillatory form with a variance that grows quadratically with time. For recent comprehensive reviews, see Kempe \cite{Ke}, Tregenna {\it et al.} \cite{TFMK}, Ambainis \cite{Am}.

A unitary matrix corresponding to the evolution of the QW is usually deterministic and independent of the time step. In this letter we study a random unitary matrix case motivated by numerical results of Ribeiro {\it et al.} \cite{RMM} and Mackay {\it et al.} \cite{MBSS}. We obtain a classical random walk from the QW by introducing an independent random fluctuation at each time step and performing an ensemble average in a rigorous way based on a path integral approach. The time evolution of the QW is given by the following infinite random unitary matrices $\{ U_n : n=1,2, \ldots \}$:
\begin{eqnarray*}
U_n=
\left[
\begin{array}{cc}
a_n & b_n \\
c_n & d_n
\end{array}
\right],
\end{eqnarray*}
\par\noindent
where $a_n,b_n,c_n,d_n \in {\bf C}$ and ${\bf C}$ is a set of complex numbers. The subscript $n$ indicates the time step. The unitarity of $U_n$ gives 
\begin{eqnarray}
|a_n|^2 + |c_n|^2 =|b_n|^2 + |d_n|^2 =1, \> a_n \overline{c_n} + b_n \overline{d_n}=0, \> c_n= - \triangle_n \overline{b_n}, \> d_n= \triangle_n \overline{a_n},\label{eqn:seisitu}
\end{eqnarray}
where $\overline{z}$ is the complex conjugate of $z \in {\bf C}$ and $\triangle_n = \det U_n = a_n d_n - b_n c_n$ with $|\triangle_n|=1.$ Put $w_n = (a_n, b_n, c_n, d_n).$ Let $\{ w_n : n=1,2, \ldots \}$ be independent and identically distributed (or i.i.d. for short) on some space, (for example, $[0, 2 \pi))$ with 
\begin{eqnarray}
&& 
E(|a_1|^2)=E(|b_1|^2)=1/2, 
\label{eqn:poki}
\\
&&
E(a_1 \overline{c_1})= 0.
\label{eqn:kuki}
\end{eqnarray}
Remark that Eq.(\ref{eqn:poki}) implies $E(|c_1|^2)=E(|d_1|^2)=1/2,$ and Eq.(\ref{eqn:kuki}) gives $E(b_1 \overline{d_1})=0$ by using Eq.(\ref{eqn:seisitu}). The set of initial qubit states for the QW is given by 
\[
\Phi = \left\{ \varphi =
{}^t [\alpha,\beta] \in {\bf C}^2 :
|\alpha|^2 + |\beta|^2 =1
\right\},
\]
where $t$ means the transposed operator. Moreover we assume that $\{w_n : n =1,2, \ldots \}$ and $\{\alpha, \beta \}$ are independent. We call the above process {\it disordered QW} in this letter. Let ${\bf R}$ be the set of real numbers. Then we consider the following two cases:

Case I: $a_n,b_n,c_n,d_n \in {\bf R} \> (n=1,2, \ldots )$ and $E(\alpha \overline{\beta}+\overline{\alpha} \beta) =0$.

Case II: $E(|\alpha|^2)=1/2$ and $E(\alpha \overline{\beta})=0$.

\par
We should note that Case I corresponds to an example given by Ribeiro {\it et al.} \cite{RMM} and Case II corresponds to an example given by Mackay {\it et al.} \cite{MBSS}, respectively. The numerical simulations by these two groups suggest that the probability distribution of the disordered QW converges to a binomial distribution by averaging over many trials. So the main purpose of the letter is to prove the above numerical results by using a path integral approach, which has been used in \cite{KNS, K001, K002}, for example. In fact, our theorem (see Theorem 1) shows that the expectation of the probability distribution for the disordered QW becomes the probability distribution of a classical symmetric random walk. However our result does not treat a crossover from quantum to classical walks. Concerning the crossover, see \cite{BCA001, BCA002, BCA003, KT001, KT002, KT003, SBBH}, for specific examples.

The rest of the letter is organized as follows. Section 2 gives the definition of the disordered QW. In Section 3, we prove Theorem 1. Section 4 is devoted to examples given by \cite{RMM, MBSS, SBBH}.

\setcounter{equation}{0}
\section{Definition of Disordered QW}
First we divide $U_n$ into two matrices:
\begin{eqnarray*}
P_n =
\left[
\begin{array}{cc}
a_n & b_n \\
0 & 0 
\end{array}
\right], 
\quad
Q_n=
\left[
\begin{array}{cc}
0 & 0 \\
c_n & d_n 
\end{array}
\right]
\end{eqnarray*}
with $U_n=P_n+Q_n$. The important point is that $P_n$ (resp. $Q_n$) represents that the particle moves to the left (resp. right) at each time step $n$. By using $P_n$ and $Q_n$, we define the dynamics of the disordered QW on the line. To do so, we define the $(4N+2) \times (4N+2)$ matrix $\overline{U}_n$ by 
\begin{eqnarray*}
\overline{U}_n =
\left[
\begin{array}{ccccccc}
0 & P_n & 0 & \dots & \dots & 0 & Q_n \\
Q_n & 0 & P_n & 0 & \dots & \dots & 0 \\
0 & Q_n & 0 & P_n & 0 &\dots &0\\
\vdots & \ddots & \ddots & \ddots & \ddots & \ddots & \vdots \\
0 & \dots & 0 & Q_n & 0 & P_n & 0\\
0 & \dots & \dots & 0 & Q_n & 0 & P_n\\
P_n & 0 & \dots & \dots & 0 & Q_n & 0
\end{array}
\right],
\end{eqnarray*}
where $0=O_2$ is the $2 \times 2$ zero matrix. Note that $P_n$ and $Q_n$ satisfy
\begin{eqnarray*}
P_nP_n ^*+Q_nQ_n ^* = P_n ^* P_n + Q_n ^* Q_n =I_2
, \quad P_n Q_n ^*=Q_n P_n ^*=Q_n ^* P_n =P_n ^* Q_n =O_2,
\end{eqnarray*}
where $*$ means the adjoint operator and $I_2$ is the $2 \times 2$ unit matrix. The above relations imply that $\overline{U}_n$ becomes also a unitary matrix. Let 
\begin{eqnarray*}
\Psi_k ^{(n)} (\varphi) = 
\left[
\begin{array}{cc}
\Psi_{L,k} ^{(n)} (\varphi) \\
\Psi_{R,k} ^{(n)} (\varphi)   
\end{array}
\right] 
\in {\bf C}^2
\end{eqnarray*}
be a two component vector of amplitudes of the particle being at site $k$ and at time $n$ with the chirality being left (upper component) and right (lower component), and let
\begin{eqnarray*}
\Psi^{(n)} (\varphi) = {}^t [\Psi_{-N} ^{(n)} (\varphi), \Psi_{-(N-1)} ^{(n)} (\varphi), \ldots, \Psi_{N} ^{(n)} (\varphi)]
\end{eqnarray*}
be the state at time $n$. An initial state for the system is given by 
\[
\Psi^{(0)} (\varphi) = {}^t[\overbrace{0, \ldots , 0,}^N \varphi, \overbrace{0, \ldots, 0}^N ] \in {\bf C}^{2N+1},
\]
where $0={}^t[0,0]$ and $\varphi = {}^t [\alpha,\beta]$. The following equation defines the time evolution of the disordered QW: for $-N \le k \le N$ and $1 \le n < N$, 
\[ 
(\Psi^{(n)} (\varphi))_k=(\overline{U}_n \Psi^{(n-1)} (\varphi))_k = Q_n \Psi_{k-1} ^{(n-1)} (\varphi) + P_n 
\Psi_{k+1} ^{(n-1)} (\varphi),  
\]
For an initial state $\overline{\varphi}={}^t[\overbrace{0,\ldots,0}^N, \varphi , \overbrace{0,\ldots,0}^N]$, we have 
\begin{eqnarray*}
&& \overline{U}_1 \overline{\varphi} = {}^t[\overbrace{0,\dots , 0}^{N-1},P_1 \varphi,0,Q_1 \varphi,\overbrace{0,\dots,0}^{N-1}], 
\\
&& \overline{U}_2 \overline{U}_1 
\overline{\varphi} ={}^t[\overbrace{0,\ldots,0}^{N-2},
P_2 P_1 \varphi,0,(P_2 Q_1 + Q_2 P_1 )\varphi,0,Q_2 Q_1 \varphi,\overbrace{0, \ldots,0}^{N-2}].
\end{eqnarray*}
Let $X_n ^{\varphi}$ denote the disordered QW at time $n$ starting from $\varphi \in \Phi.$ By using $\Psi_k ^{(n)} (\varphi)$, the probability of $\{ X_n ^{\varphi} = k \}$ is defined by
\[
P(X_n ^{\varphi} = k) 
= \| \Psi_k ^{(n)} (\varphi) \|^2
= | \Psi_{L,k} ^{(n)} (\varphi) |^2 + | \Psi_{R,k} ^{(n)} (\varphi) |^2.
\]
Remark that $\{ X_n ^{\varphi} = k \}$ is an event generated by $\{w_i:i=1,2, \ldots ,n \}$ and $\{\alpha, \beta \}.$ The unitarity of $\overline{U}_m (m=1,2, \ldots, n)$ ensures
\[
\sum_{k=-n} ^{n} P(X_n ^{\varphi} = k) 
= \| \overline{U}_n \overline{U}_{n-1} \cdots \overline{U}_1 
\overline{\varphi} \|^2
= \| \overline{\varphi} \|^2
= |\alpha|^2 + |\beta|^2
= 1,
\]
for any $1 \le n \le N$. That is, the amplitude always defines a probability distribution for the location. 

\setcounter{equation}{0}
\section{Main Result}
\hspace*{1em}
We now give a combinatorial expression of the probability distribution of
 the disordered QW. For fixed $l$ and $m$ with $l+m=n$ and $-l+m=k$, we introduce
\[
\Xi_n (l,m)= \sum_{l_j, m_j \ge 0: l_1+ \cdots +l_n=l, m_1+ \cdots +m_n=m} P^{l_n}Q^{m_n}P^{l_{n-1}}Q^{m_{n-1}} \cdots P^{l_1}Q^{m_1}.
\]
We should note that
\[
\Psi_k ^{(n)} (\varphi) = \Xi_n (l,m) \varphi.
\]
For example, in the case of $P(X_4 ^{\varphi} = 0)$, we have 
\begin{eqnarray*}
\Xi_4 (2,2) 
&=&
P_4 Q_3 Q_2 P_1 + Q_4 P_3 P_2 Q_1
\\
&&
+ P_4 P_3 Q_2 Q_1 + P_4 Q_3 P_2 Q_1 +
Q_4 P_3 Q_2 P_1 + Q_4 Q_3 P_2 P_1. 
\end{eqnarray*}
Here we introduce the useful random matrices to compute $\Xi_n (l,m)$:
\[
R_n =
\left[
\begin{array}{cc}
c_n  & d_n \\
0 & 0 
\end{array}
\right], 
\quad
S_n =
\left[
\begin{array}{cc}
0 & 0 \\
a_n & b_n 
\end{array}
\right].
\]
In general, we obtain the following table of products of the matrices $P_j,Q_j,R_j$ and $S_j$ ($j=1,2, \ldots)$: for any $m, n \ge 1,$ 
\par
\
\par
\begin{center}
\begin{tabular}{c|cccc}
  & $P_n$ & $Q_n$ & $R_n$ & $S_n$  \\ \hline
$P_m$ & $a_mP_n$ & $b_mR_n$ & $a_mR_n$ & $b_mP_n$  \\
$Q_m$ & $c_mS_n$ & $d_mQ_n$ & $c_mQ_n$ & $d_mS_n$ \\
$R_m$ & $c_mP_n$ & $d_mR_n$ & $c_mR_n$ & $d_mP_n$ \\
$S_m$ & $a_mS_n$ & $b_mQ_n$ & $a_mQ_n$ & $b_mS_n$ 
\end{tabular}
\end{center}
where $P_m Q_n = b_m R_n$, for example. From this table, we obtain
\begin{eqnarray*}
\Xi_4 (2,2) 
&=&
b_4 d_3 c_2 P_1 + c_4 a_3 b_2 Q_1
\\
&&
+ (a_4 b_3 d_2 + b_4 c_3 b_2) R_1 +
(c_4 b_3 c_2 + d_4 c_3 a_2) S_1. 
\end{eqnarray*}
\par
We should note that $P_1, Q_1, R_1$ and $S_1$ form an orthonormal basis of the vector space of complex $2 \times 2$ matrices with respect to the trace inner product $\langle A | B \rangle = $ tr$(A^{\ast}B)$. So $\Xi_n (l,m)$ has the following form:
\begin{eqnarray}
\Xi_n (l,m) = p_n (l,m) P_1 + q_n (l,m) Q_1 + r_n (l,m) R_1 + s_n (l,m) S_1.
\label{eqn:yuki}
\end{eqnarray}
In general, explicit forms of $p_n (l,m), q_n (l,m), r_n (l,m)$ and $s_n (l,m)$ are complicated. From Eq. (\ref{eqn:yuki}), we have
\begin{eqnarray*}
\Xi_n (l,m) 
=
\left[
\begin{array}{cc}
p_n (l,m) a_1 + r_n (l,m) c_1 & 
p_n (l,m) b_1 + r_n (l,m) d_1 \\
q_n (l,m) c_1 + s_n (l,m) a_1 & 
q_n (l,m) d_1 + s_n (l,m) b_1
\end{array}
\right].
\label{eqn:aki}
\end{eqnarray*}
Remark that $p_n (l,m), q_n (l,m), r_n (l,m), s_n (l,m)$ depend only on $\{ w_i : i=2,3, \ldots n \}$, so they are independent of $w_1$, where $w_i = (a_i, b_i, c_i, d_i).$ Moreover it should be noted that $\{w_i : i =1,2, \ldots \}$ and $\{ \alpha, \beta \}$ are independent. Therefore we obtain
\begin{eqnarray*}
P(X_n ^{\varphi} =k)
&=&
|| \Xi_n (l,m) \varphi ||^2
\\
&=& 
\left\{ 
|p_n (l,m) a_1 + r_n (l,m) c_1|^2 + 
|q_n (l,m) c_1 + s_n (l,m) a_1|^2 \right\} | \alpha |^2
\\
&&
+
\left\{
|p_n (l,m) b_1 + r_n (l,m) d_1|^2 + 
|q_n (l,m) d_1 + s_n (l,m) b_1|^2 \right\} | \beta |^2
\\
&&
+
\left\{ 
\overline{(p_n (l,m) a_1 + r_n (l,m) c_1)}(p_n (l,m) b_1 + r_n (l,m) d_1) 
\right.
\\
&&
\left.
+ 
\overline{(q_n (l,m) c_1 + s_n (l,m) a_1)}(q_n (l,m) d_1 + s_n (l,m) b_1) 
 \right\} 
\overline{\alpha} \beta 
\\
&&
+
\left\{ 
(p_n (l,m) a_1 + r_n (l,m) c_1) \overline{(p_n (l,m) b_1 + r_n (l,m) d_1)} 
\right.
\\
&&
\left.
+ 
(q_n (l,m) c_1 + s_n (l,m) a_1) \overline{(q_n (l,m) d_1 + s_n (l,m) b_1)} 
\right\} 
\alpha \overline{\beta} 
\\
&=&
C_1 | \alpha |^2+ C_2 | \beta |^2 
+ C_3 \overline{\alpha} \beta + C_4 \alpha \overline{\beta}. 
\end{eqnarray*}
Then we see that 
\begin{eqnarray*}
E(C_1 | \alpha |^2)
=
{1 \over 2} E \bigl[|p_n (l,m)|^2  + |s_n (l,m)|^2 + |q_n (l,m)|^2  + |r_n (l,m)|^2 \bigr] E(| \alpha |^2),
\end{eqnarray*}
since $\{ p_n (l,m), q_n (l,m), r_n (l,m), s_n (l,m)\}$ and $w_1$ are independent, $\{w_i : i =1,2, \ldots \}$ and $\{ \alpha, \beta \}$ are independent, Eqs.(\ref{eqn:poki}) and (\ref{eqn:kuki}). Similarly, we have 
\begin{eqnarray*}
E(C_2 | \beta |^2)
=
{1 \over 2} E \bigl[ |p_n (l,m)|^2  + |s_n (l,m)|^2  
+  |q_n (l,m)|^2  + |r_n (l,m)|^2 \bigr] E(| \beta |^2).
\end{eqnarray*}
Case I: First we note that $C_4 = \overline{C_3}$. So $a_n, b_n, c_n, d_n \in {\bf R}$ gives $C_3 = C_4$. Then the condition $\alpha \overline{\beta}+\overline{\alpha} \beta =0 $ implies $C_3 \overline{\alpha} \beta + C_4 \alpha \overline{\beta} =0$. Therefore we obtain
\begin{eqnarray*}
&&
E(P(X_n ^{\varphi} =k))
\nonumber
\\
&&
=
E(|| \Xi_n (l,m) \varphi ||^2)
\\
&&
=  
{1 \over 2} 
E \bigl[ |p_n (l,m)|^2  + |s_n (l,m)|^2 + |q_n (l,m)|^2  + |r_n (l,m)|^2 
\bigr] \>
E(| \alpha |^2) 
\\
&&
+ 
{1 \over 2} 
E \bigl[ |p_n (l,m)|^2  + |s_n (l,m)|^2 + |q_n (l,m)|^2  + |r_n (l,m)|^2 
\bigr] \>
E(| \beta |^2)
\\
&&
=
{1 \over 2} \left\{  
E(|p_n (l,m)|^2) + E(|q_n(l,m)|^2) + E(|r_n(l,m)|^2) + E(|s_n(l,m)|^2) 
\right\},
\label{eqn:kazue}
\end{eqnarray*}
since $|\alpha|^2+|\beta|^2=1$.

\par\noindent
Case II: $E(\alpha \overline{\beta})=E(\overline{\alpha} \beta)=0$ gives the above same conclusion in a similar fashion.

To get a feel for the following main result, we will look at a concrete case of $n=4$ with $l=m=2$.
\begin{eqnarray*}
&& 
E( P(X_4 ^{\varphi} =0))
\\
&=&
E(|| \Xi_4 (2,2) \varphi ||^2)
\\
&=& {1 \over 2} \left\{
E(|p_4(2,2)|^2) + E(|q_4(2,2)|^2) + E(|r_4(2,2)|^2) + E(|s_4(2,2)|^2)
\right\}
\\
&=&
{1 \over 2} \left\{
E(|b_4|^2) E(|d_3|^2) E(|c_2|^2) 
+ E(|c_4|^2) E(|a_3|^2) E(|b_2|^2) 
\right.
\\
&&
+ E(|a_4|^2) E(|b_3|^2) E(|d_2|^2)
+ E(|b_4|^2) E(|c_3|^2) E(|b_2|^2)
\\
&&
+ E(a_4 \overline{b_4}) E(b_3 \overline{c_3}) E(d_2 \overline{b_2})
+ E(\overline{a_4} b_4) E(\overline{b_3} c_3) E(\overline{d_2} b_2)
\\
&&
+ E(|c_4|^2) E(|b_3|^2) E(|c_2|^2)
+ E(|d_4|^2) E(|c_3|^2) E(|a_2|^2)
\\
&&
\left.
+ E(c_4 \overline{d_4}) E(b_3 \overline{c_3}) E(c_2 \overline{a_2})
+ E(\overline{c_4} d_4) E(\overline{b_3} c_3) E(\overline{c_2} a_2)
\right\}
\\
&=&
{1 \over 16} + {1 \over 16} + {2 \over 16} + {2 \over 16} = {1 \over 2^4} \> {4 \choose 2}, 
\end{eqnarray*}
since $E(|a_1|^2)=E(|b_1|^2)=E(|c_1|^2)=E(|d_1|^2)=1/2,$ and $E(a_1 \overline{c_1})=E(b_1 \overline{d_1})=0.$ This result corresponds to $P(Y^o _4 = 0) = {4 \choose 2}/2^4,$ where  $Y^o _n$ denotes the classical symmetric random walk starting from the origin. We generalize the above example as follows:

\begin{thm}
\label{thm:thm1} We assume that a disordered QW starting from $\varphi$ satisfies Case I or Case II. For $n=0,1,2, \ldots, $ and $k=-n, -(n-1), \ldots, n-1,n,$ with $n+k$ is even, we have
\begin{eqnarray*}
E(P(X^{\varphi} _n=k)) = P(Y^o _n = k) = 
{1 \over 2^n} \> {n \choose (n+k)/2} 
\end{eqnarray*}
\end{thm}

\noindent
Proof. For $n=0,1,2, \ldots, $ and $k=-n, -(n-1), \ldots, n-1,n,$ with $n+k$ is even, let $l=(n-k)/2, \> m=(n+k)/2$ and $M={n \choose (n+k)/2}$. In general, as in the case of $n=4$ with $l=m=2$, we have 
\begin{eqnarray*}
P(X^{\varphi} _n=k)
&=&
E(|| \Xi_n (l,m) \varphi ||^2)
\\ 
&=& {1 \over 2} \sum_{j=1}^M E(|u_n ^{(j)}|^2) E(|u_{n-1} ^{(j)}|^2) 
\cdots E(|u_2 ^{(j)}|^2) + E(R(w_1, w_2, \ldots, w_n)), 
\end{eqnarray*}
where $u_i ^{(j)} \in \{ a_i, b_i, c_i, d_i \}$ for any $i=2,3, \ldots, n, \> j=1,2, \ldots, M$ and $w_k =(a_k, b_k, c_k, d_k)$ for any $k=1,2, \ldots, n.$ Then $E(|a_1|^2)=E(|b_1|^2)=E(|c_1|^2)=E(|d_1|^2)=1/2$ gives
\begin{eqnarray*}
E(P(X^{\varphi} _n=k)) = {M \over 2^n}+ E(R(w_1, w_2, \ldots, w_n)). 
\end{eqnarray*}
To get the desired conclusion, it suffices to show that $E (R (w_1, w_2, \ldots, w_n)) = 0$. When there are more than two terms in $p_n (l,m)$ or $s_n (l,m)$, we have to consider two cases: there is a $k$ such that 
\begin{eqnarray*}
\cdots P_{k+1} P_k P_{k-1} \cdots P_1, \qquad  \cdots Q_{k+1} P_k P_{k-1} \cdots P_1, 
\end{eqnarray*}
For example, $s_4 (2,2)=d_4 c_3 a_2 + c_4 b_3 c_2$ case is $k=1$, since 
\begin{eqnarray*}
Q_4 Q_3 P_2 P_1, \qquad Q_4 P_3 Q_2 P_1.
\end{eqnarray*}
Then the corresponding $p_n (l,m)$ or $s_n (l,m)$ are given by
\begin{eqnarray*}
\cdots a_{k+1} a_k a_{k-1} \cdots a_2, \qquad  \cdots c_{k+1} a_k a_{k-1} \cdots a_2.
\end{eqnarray*}
Therefore we have
\begin{eqnarray*}
&&
E \> \bigl[ \overline{(\cdots a_{k+1} a_k a_{k-1} \cdots a_2)}(\cdots c_{k+1} a_k a_{k-1} \cdots a_2) \bigr]
\\
&=&  
E ( \cdots ) \> E(\overline{a_{k+1}} c_{k+1}) \> E(|a_k|^2) \> 
E ( |a_{k-1}|^2 ) \cdots E(|a_2|^2) =0,
\end{eqnarray*}
since $E(\overline{a_1}c_1)=0$. In other cases, a similar argument holds. So the proof is complete.

The above theorem implies that the expectation of the probability distribution for the disordered QW is nothing but the probability distribution of classical symmetric random walk. 

\setcounter{equation}{0}
\section{Examples}
\hspace*{1em}
The first example was introduced and studied by Ribeiro {\it et al.} [4]. This corresponds to Case I:
\begin{eqnarray*}
U_n =
\left[
\begin{array}{cc}
\cos (\theta_n) & \sin (\theta_n) \\
\sin (\theta_n) & - \cos (\theta_n)
\end{array}
\right],
\end{eqnarray*}
where $\{ \theta_n : n=1,2, \ldots \}$ are i.i.d. on $[0, 2 \pi)$ with 
\begin{eqnarray*}
E(\cos^2 (\theta_1))=E(\sin^2 (\theta_1))=1/2, \quad 
E(\cos (\theta_1) \sin (\theta_1))=0.
\end{eqnarray*}
Here we give two concrete cases:

(i) $\theta_1$ is the uniform distribution on $[0, 2 \pi)$, 

(ii) $P(\theta_1 = \xi)=P(\theta_1 = \pi/2 + \xi) =1/2$ for some $\xi \in [0, \pi)$.
\par\noindent
For an initial qubit state, we choose a non-random $\varphi = {}^t[\alpha, \beta] \in \Psi$ with $|\alpha|= |\beta|=1/\sqrt{2},$ and $\alpha \overline{\beta} + \overline{\alpha} \beta =0$.

The second example was given by Mackay {\it et al.} [5]. This corresponds to Case II: 
\begin{eqnarray*}
U_n ={1 \over \sqrt{2}}
\left[
\begin{array}{cc}
1 & e^{i \theta_n} \\
e^{-i \theta_n} & - 1
\end{array}
\right],
\end{eqnarray*}
where $\{ \theta_n : n=1,2, \ldots \}$ are i.i.d. on $[0, 2 \pi)$ with 
\begin{eqnarray*}
E(\cos (\theta_1))=E(\sin (\theta_1))=0.
\end{eqnarray*}
Moreover an initial qubit state is chosen as $\varphi = {}^t[\alpha, \beta] \in \Psi$ with $E(|\alpha|^2)=1/2$ and $E(\alpha \overline{\beta})=0$. For example, both $\theta_1$ and $\theta_{\ast}$ are uniform distributions on $[0, 2 \pi),$ and they are independent each other. Let $\alpha = \cos (\theta_{\ast})$ and $\beta = \cos (\theta_{\ast})$. So the above conditions hold.

Finally we discuss an example studied by Shapira {\it et al.} \cite{SBBH}. Let $W_n = \{X_n, Y_n, Z_n \},$ where $\{W_n:n=1,2, \ldots \}$ is i.i.d., and $\{X_n, Y_n, Z_n \}$ is also i.i.d., moreover $X_n$ is the normal distribution with the mean 0 and the variance $\sigma^2$ with $\sigma >0.$ Their case is 
\begin{eqnarray*}
U_n ={1 \over \sqrt{2}}
\left[
\begin{array}{cc}
1 & 1 \\
1 & - 1
\end{array}
\right] \times V_n,
\end{eqnarray*}
where
\begin{eqnarray*}
V_n ={1 \over \sqrt{2}}
\left[
\begin{array}{cc}
\cos (R_n) + i Z_n {\sin (R_n) \over R_n} &  (Y_n + i X_n) {\sin (R_n) \over R_n} \\
(- Y_n + i X_n) {\sin (R_n) \over R_n}  & \cos (R_n) - i Z_n {\sin (R_n) \over R_n}
\end{array}
\right],
\end{eqnarray*}
and $R_n = \sqrt{X_n ^2 + Y_n ^2 + Z_n ^2}.$ Then a direct computation implies 
\begin{eqnarray}
&& 
E(|a_1|^2)=E(|b_1|^2)=1/2, 
\nonumber
\\
&&
E(a_1 \overline{c_1})= {1 \over 6}+{1 \over 3}(1- 4 \sigma^2) e^{-2 \sigma^2}
\equiv \mu (\sigma).
\label{eqn:pinki}
\end{eqnarray}
We see that Eq. (\ref{eqn:pinki}) implies $0< \mu (\sqrt{3}/2)=0.0179 \ldots \le \mu (\sigma) \le 1/2 (= \lim_{\sigma \downarrow 0} \mu (\sigma))$ for any $\sigma >0.$ Note that the limit $\sigma \downarrow 0$ corresponds to the Hadamard walk case. So their example does not satisfy our condition Eq. (\ref{eqn:kuki}).

\appendix 


\end{document}